\documentclass{article}
\usepackage{graphicx}
\begin{document}                                 
\noindent {\bf Crab pulsar spectrum: a non-extensive approach }
\vskip1.5cm
\noindent {\bf Ashok Razdan}

\noindent {\bf Astrophysical Sciences Division }

\noindent {\bf Bhabha Atomic Research Centre }

\noindent {\bf Trombay, Mumbai- 400085 }
\vskip 0.5cm
Key words : Crab pulsar, non-extensive statistical physics, $\gamma$-ray astronomy, Inverse Compton scattering  
\vskip 0.5cm
\noindent {\bf Abstract :}

Non extensive statistical physics has been applied to various problems in physics including astrophysics. In
this paper we explore the possibility of using non extensive approach to explain the recently observed pulsed
$\gamma$-ray from Crab pulsar above 100 GeV observed by VERITAS $\gamma$-ray telescope.
\vskip 0.5cm
\noindent{\bf Motivation :}

The recent detection of pulsed $\gamma$-ray above 100 GeV from Crab pulsar [1] by VERITAS $\gamma$-ray telescope
can not be explained by standard pulsar models and data has been parametrized by
broken power law 
instead of standard model of  power law with exponential cutoff. 
CRAB  is a supernova remenent which  emits in radio, optical, x-ray and soft $\gamma$-ray wavelengths.  
In the present paper  we explore the possibility of
using non extensive form of power law with exponential function to parametize  the CRAB pulsar spectrum.
Non-extensive  statistical physics  has been  used to explain
anomalous results  observed in  various  problems  
in physics [2-8].
Non-extensive features get manifested in those systems 
which have long range forces, long memory effects,inhomogeneous systems or in those systems which 
evolve in (non Euclidean like space-time) fractal space time [8 and reference therein]. 

Non-extensive statistical physics has been applied to various astrophysics problems also.
It has 
been shown that galaxy distribution [9] follows non extensive statistical physics.
Tsallis et. al. [10]  studied links between various astronomical phenomena and non extensive
statistical physics.
Luis et. al. [11] used non-extensive approach to study galactic halos  
of self gravitating systems. Due to long range nature of gravity, self gravitating systems follow non extensive statistics  and
Juilin [12] obtained non extensive form of Maxwellian distribution which is 
applicable to astrophysical environments. Chavanis et. al.[13] used non extensive approach   
to investigate non linear stability of stars and galaxies.
Armondo et. al. [14] has shown that the distribution of  temperature
fluctuations in cosmic microwave background 
follows non extensive form of Gaussian distribution to  the confidence level of 99 $\%$.
Using non extensive form of velocity distributions, Chavalho et.al. [15] studied rotational velocities
of more than 1600 F and G dwarf stars.

Non extensive statistical physics has been used to derive the density distribution of large scale astrophysical structures
and derived profiles have perfect match [16] with the profiles obtained from simulated dark matter and hot plasma distributions.
Non extensive approach has also been used to study density profile of simulated galaxy sized dark matter haloes [17].
It has been shown that solar carona dynamics has non Gaussian and non extensive [18,19] statistical character 
and  anomalous diffusion of electron and proton in solar core plasma can be explained 
by non extensive approach. The study of scale dependence of intermittent flows in astrophysical plasma
turbulence which arises due to long range interactions [20]  can be explained using non extensive approach.
V.M.Vasylivnas [21] used Kappa distributions to account for high energy 
tail velocity distributions of electrons and ions in space plasma. The link between Kappa distributions and velocity distributions
of non extensive nature was established by Leubner [22] resulting in to the theoretical justification of Kappa distributions in
the astrophysical plasma. One of the important features of astrophysical plasma is the non Maxwellian core-halo electron
and ion velocity distributions and non-extensive approach [23]  has been used to explain the results.

\noindent{ \bf non-extensive approach:}

Standard Statistical Physics ( Boltzmann-Gibbs thermostatistics) holds as long as thermodynamic
extensivity  (additivity) holds i.e. when
(a) effective microscopic interactions are short range and
(b) systems evolve in
Euclidean like space-time ( a continuous and sufficiently differentiable)
For two systems A and B entropy is additive
\begin{equation}
 S(A+B)= S(A)+S(B)
\end{equation}
Boltzmann-Gibbs(BG)entropy is additive and extensive.
BG approach fails 
(a) in systems with long range forces or long memory effects
(b) or  if systems evolve in non Euclidean space-time( i.e. fractals or multifractals).
Such systems which do not follow Boltzmann-Gibbs approach 
are called as non-extensive systems [ 7 and references therein].
For two systems A and B in non extensive approach
\begin{equation}
 S(A+B)= S(A)+S(B)+(1-q)S(A) S(B)
\end{equation}
where q is non extensive index.
Non-extensive statistics is based on two postulates of entropy and internal energy.
Non-extensive entropy [7] is defined as
\begin{equation}
S_q  = \frac{ 1- \sum_i P_i^{q}}{q -1}
\end{equation}
In the limit of q $\rightarrow$ 0 , entropy is given as
\begin{equation}
S= -k  p_{i} ln {p_i}
\end{equation}
which is Boltzmann-Gibbs Shannon entropy.
In non-extensive approach exponential function is written as
\begin{equation}
 e_q ^{\pm x}  = [1 \pm (1-q)x]^ \frac{1}{1-q}
\end{equation}

\noindent{ \bf Crab Pulsar results : }

In  recent discovery of 100 GeV energy from the CRAB, curvature radiation process 
has been ruled out as a possible mechanism.
It has been argued that inverse Compton process may be  responsible for high energy emission from CRAB pulsar. 
Inverse Compton scattering  is very important process for $\gamma$-ray production in astrophysics. In this process low
energy photon interaction with relativistic electrons leads to boosting of low photon energies to high energies
in various astrophysical environments like pulsars, active galactic nuclei, supernova remenents , cluster of galaxies etc.
Lyutikov et. al. [24,25] have shown that inverse Compton effect is
the main emission mechanism for CRAB spectrum with non-exponential cutoff. Lefa et. al. [26] have discussed shape of radiation
distributions due to inverse Compton in various scenario's including Klein-Nishina regime. In this regime upscattered photons
pick up almost all the energy of the relativistic electrons.

The study of electromagnetic radiation from CRAB pulsar has attracted attention in all energy ranges. It began with
COMPTEL (Imaging Compton telescope) and EGRET (Energetic $\gamma$-rays  Experiment Telescpe) [27] observation.
COMPTEL detected signal from CRAB pulsar from 1 to 30 Mev range.
The EGRET  measured $\gamma$-ray spectrum of CRAB pulsar from 30 Mev  up to 5 GeV energy which followed 
power law spectrum $F(E) \propto E^{-\alpha}$ where $\alpha$=2.022 $\pm$0.014 where F is flux. Pulsar emission energy spectrum are described by a generalized form
\begin{equation}
 F(E) = A (\frac{E}{E_c})^{-\alpha} exp [-(\frac{E}{E_c})^{\beta}] 
\end{equation}
where A  is the normalized constant , $E_c$ is the cutoff energy and $\beta$ describes the
steepness of the cutoff. However, up to 5 GeV no cutoff in energy spectrum of CRAB pulsar was observed  by EGRET. The observation of
CRAB pulsar by MAGIC telescope resulted in  signal above 25 GeV and measured flux was several times lower than the flux measured by EGRET 
which signalled the appearance of cutoff. A joint fit of data obtained in different energy ranges like 1-30 MeV  from COMPTEL, 30 MeV to
10 GeV from EGRET data and MAGIC data above 25 GeV was observed to follow  power law with exponential cutoff.
It was further reported  that  there is similarity in EGRET data and Fermi-LAT data on CRAB pulsar. VERITAS observations
on CRAB pulsar in 100 GeV region  provided additional data to study its energy spectrum. Joint fitting of Fermi-LAT data, MAGIC data and
VERITAS data by power law with exponential cutoff produced high value of $\chi^{2}$ =66 for 16 degrees of freedom. Broken power law fit to 
the data has $\chi^{2}$ valaue of 13.5 for 15 degress of freedom. Now we explore the possiblity of using non extensive form of power law
to fit to CRAB pulsar data.

Non extensive form of equation (6) can be obtained by Using equation (5) in equation (6), we have 
\begin{equation}
F(E)= \frac{dN}{dE} = A  (\frac{E}{E_0})^{-\alpha} [1-(1-q)(\frac{E}{E_{c}})^{\beta}]^{\frac{1}{1-q}}
\end{equation}
We have used equation(7) to
fit to the data of CRAB spectrum over the whole energy range (COMPTEL, EGRET, MAGIC and  VERITAS) data as shown in figure 1.
We have used gnuplot to fit equation (7) by initially fixing q=1.2. The CRAB pulsar data over whole energy range
has been fiited by allowing  A, $E_0$, $\alpha$, $E_c$, $\beta$ and q as variables. The final values 
are 
$E_c$ = 551.562 MeV,
A= 3.67487e-05,
$\alpha$ = -0.413736,
$E_0$= 1.43185 MeV,
$\beta$ = 0.57564 
and q = 1.2 .
For this  case of fitting,  the degrees of freedom are 7,
the rms of residuals is 0.9249
and the variance of residuals (reduced $\chi^{2}$ ) is equal to 0.85544.

Pulsars are rotating magnetized neutron stars with corotating magnetosphere where charged particles are accelerated to
relativistic energies. 
Lyutikov [24,25] have shown that in the case of CRAB pulsar,
scattering may happen from dense secondary plasma in deep Klein-Nishina regime 
where inverse scattered photon may provide direct measurement of particle distribution within the magnetosphere.
The spectral energy distribution of observed photons is dependent on the energy distribution of relativistic  electrons and
it may be possible that velocities of the relativistic  electrons may be following non-extensive Maxwallian distribution.
 
Various anomalous results observed in astrophysical plasma have been resolved by using non extensive approach.
Ferro et. al.[28]
studied non extensive resonant reaction rates in astrophysical plasma's in detail. 
Soares et. al.[29] discusses the superiority of non-extensive Maxwell velocity distribution
over  standard  Maxwell velocity distribution for the case of rotational velocities of
stars in Pleiadus open clusters.
Freites et. al.[30] studied evolution of F and
G stars and angular momentum loss by magnetic stellar wind with non extensive statistical physics.
In an interesting   studies Saxena et.al.[31]
derived astrophysical thermonuclear functions both for Boltzmann Gibbs statistics and for non extensive statistics and
it was observed the cut off
at higher energies can be explained  better by non extensive approach.
\begin{figure}
\includegraphics[angle=270,width=7.cm]{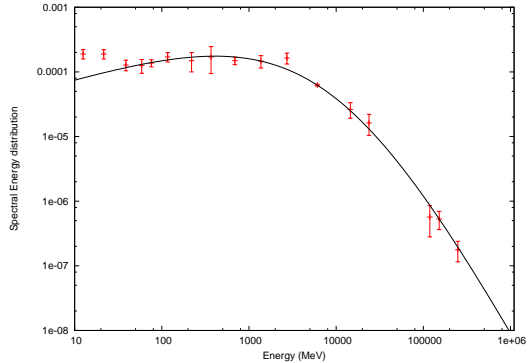}
\caption{ Parameterization of spectral energy distribution  is shown of crab pulsar data. 
The fit corresponds to equation (7) for q=1.2} 
\end{figure}

 Non-extensive power law arises due to some inhomogeneity in physical systems.
 Non-extensive distributions have been obtained using 
 Langevin equation with temperature fluctuation or multiplicative noise which are important factors in condensed matter physics.
 However in astrophysical situations which involve long range interactions, multiplicative noise or fluctuations in temperature
 may not be  factors responsible for inhomogeneity.
 In recent paper  which is of relevance  to  astrophysics, Zheng et. al.[32] has proposed  mechanism which leads to  non extensive power law
 distribution in self gravitational systems.
 In this mechanism inhomogenity arises due to inhomogenity 
 of phase space of self gravitational systems and
 and its momentum space.
 The inner friction force and momentum are important
 factors in astrophysical situations because inner friction coefficient is related to the kinetic energy of the particle.

\noindent{\bf Conclusions:}

In this paper we have shown that Crab pulsar spectrum can be explained by using non-extensive approach.  
This scenario  may arise due to non-extensive nature  of the Maxwellian distributions of the velocities of
the relativistic electrons.

\end{document}